\documentstyle[12pt]{article}
\input epsf

\begin{document}
\baselineskip=15pt
\newcommand{\x}{{\bf x}}
\newcommand{\y}{{\bf y}}
\newcommand{\z}{{\bf z}}
\newcommand{\bp}{{\bf p}}
\newcommand{\A}{{\bf A}}
\newcommand{\B}{{\bf B}}
\newcommand{\p}{\varphi}
\newcommand{\del}{\nabla}
\newcommand{\be}{\begin{equation}}
\newcommand{\ee}{\end{equation}}
\newcommand{\bq}{\begin{eqnarray}}
\newcommand{\eq}{\end{eqnarray}}
\newcommand{\ba}{\begin{eqnarray}}
\newcommand{\ea}{\end{eqnarray}}
\def\r{\nonumber\cr}
\def\hf{\textstyle{1\over2}}
\def\qr{\textstyle{1\over4}}
\def\Sc{Schr\"odinger\,}
\def\sc{Schr\"odinger\,}
\def\'{^\prime}
\def\>{\rangle}
\def\<{\langle}
\def\-{\rightarrow}
\def\dbd{\partial\over\partial}
\def\tr{{\rm tr}}
\def\hg{{\hat g}}
\def\ca{{\cal A}}
\def\pd{\partial}
\def\dl{\delta}
%%%%%%%%%%%%%%%%%%%%%%%%%%%%%%%%%%%%%%%%%%%%%%%%%%%%%%%%%%%%%%%%%%%%

\begin{titlepage}
\vskip1in
\begin{center}
{\large Topology, normalisability and the Schr\"odinger equation:
Compact QED (2+1).}
\end{center}
\vskip1in
\begin{center}
{\large David Nolland}

\vskip20pt

Department of Mathematical Sciences

University of Liverpool

Liverpool, L69 3BX, England

{\it nolland@liv.ac.uk}

\end{center}
\vskip1in
\begin{abstract}
\noindent For the special case of compact QED in (2+1) dimensions,
we calculate the non-Gaussian vacuum wave-functional to second
order in the monopole fugacity and obtain the effective photon
mass. Our method presents some hope for understanding the
connection between variational and systematic approaches to
understanding the non-perturbative wave-functional.

\end{abstract}

\end{titlepage}

\section{Introduction}

In recent years there has been a revival of interest in
Hamiltonian methods for Yang-Mills gauge theories. These studies
have for the most part fallen into two groups: one group tries to
solve the functional Schr\"odinger equation in some sort of
systematic expansion \cite{1}-\cite{9}, the other makes use of a
Gaussian ansatz for the wave-functional, and a variational
principle that minimises the vacuum energy on the Gaussian state
\cite{10}-\cite{14}. Both of these approaches have led to
interesting insights into the vacuum structure of non-abelian
gauge theories, but in both cases many important questions remain
unanswered.

Systematic methods run into a number of technical difficulties
that have been resolved in some cases but not in general. One of
the first problems is to rewrite the Hamiltonian in terms of
suitable gauge-invariant variables. This must be done in such a
way that the relevant systematic expansion is well-defined and
renormalisable. Certain expectations, such as the existence of a
derivative expansion of the wave-functional at large distances
\cite{15} are frustrated by complications arising from the
existence of topologically non-trivial classical field
configurations \cite{7,16}; some choices of gauge-invariant
variables can take them into account implicitly, or one can
perform an explicit expansion around classical solutions with
non-trivial toplogy. Dynamically generated mass parameters emerge
from the analysis in various ways \cite{7,11}.

The variational approach sidesteps many of these issues, at the
cost of restricting to a (gauge-projected) Gaussian state. This
allows many concrete calculations to be performed; on the other
hand, though it is reasonable to expect that the Gaussian state
captures the important physics, this is difficult to justify. It
would be desirable to ``elevate'' the results of the variational
method to the full non-Gaussian theory, or at least to understand
the relation between the two.

In this paper we will achieve this for the simple case of compact
QED in (2+1) dimensions, extending the analysis of \cite{11} to
the full non-Gaussian theory. This model exhibits many interesting
features such as dynamical mass generation and confinement, and
though the mechanisms for these features are special to (2+1)
dimensions, many general features of our method are more widely
applicable. Our analysis makes use of a systematic expansion in
the monopole fugacity, which is closely related to the dynamically
generated photon mass. Although the dynamically generated mass is
of the order of the monopole fugacity, to calculate it in the
non-Gaussian theory requires an analysis to second order in this
parameter. Results to higher order are obtainable in principle.

The variational principle is essentially trivial, and the meat of
the calculation is in the effective monopole dynamics, but the
former nevertheless allows us to determine the contributions of
higher $n$-point functions to the propagator and solve the
functional Schr\"odinger equation order by order. Normalisability
conditions on the wave-functional also play a role.

The ultimate aim of our investigations is to make progress towards
Hamiltonian methods for the study of QCD and Yang-Mills in (3+1)
dimensions. It is our belief that though such methods are
currently rudimentary, they can be developed to the point where
such methods can rival the explicit calculations available, for
example, in lattice gauge theory.

\section{Free QED}

To begin with consider free non-compact electrodynamics in (2+1)
dimensions. The Hamiltonian is

\be H=\hf(B^2+E_i^2), \ee where $B=\epsilon_{ij}\partial_iA_j$ and
the electric field operator is represented as

\be E_i\sim-i{\delta\over\delta A_i}. \ee If we make a Gaussian
ansatz for the vacuum wave-functional

\be \Psi[B]=\exp\{-\hf BC  B\}, \ee (here we have made use of a
matrix notation, so that $B C  B=\int d^2x d^2y B(x)C (x-y)B(y)$,
etc.) then the Schr\"odinger equation $H\Psi=E\Psi$ implies that
(in momentum space)

\be -p^2C (p)^2+1=0, \qquad E/V={1\over2}\int
{d^2p\over4\pi^2}p^2C (p).\label{fqed} \ee Here $E$ is the vacuum
energy and $V$ is the volume of space. From (\ref{fqed}) we
immediately obtain $C (p)=1/p$.

We can also derive this result by means of a variational
principle: according to the standard prescription for calculating
expectation values we have

\be \langle B^2\rangle={\int DA_i B^2\Psi[B]\Psi^*[B]\over\int
DA_i\Psi[B]\Psi^*[B]}={1\over2}\int {d^2p\over4\pi^2}{1\over C
(p)}\label{ex}, \ee while

\be \langle E_i^2\rangle={\int DA_i E_i^2\Psi[B]\Psi^*[B]\over\int
DA_i\Psi[B]\Psi^*[B]}={1\over2}\int {d^2p\over4\pi^2}{p^2C (p)}.
\label{ee}\ee Minimising the expectation value of the Hamiltonian
then gives $C (p)=1/p$ as before (and the virial theorem for
harmonic oscillators is satisfied since $\langle
E_i^2\rangle=\langle B^2\rangle$).

There is a third way of deriving this result that does not depend
on making a Gaussian ansatz. If we assume a more general form for
the vacuum wave-functional\footnote{In matrix notation ${\it C}_4
{\it B}^4=\int {\it d}^2{\it x}_1$...$ {\it d}^2{\it x}_4{\it
C}_4({\it x}_1,{\it x}_2,{\it x}_3,{\it x}_4){\it B(x}_1){\it
B(x}_2){\it B(x}_3){\it B(x}_4)$, etc.}

\be \Psi[B]=\exp\{-\hf BC  B-{1\over4!}C _4 B^4-{1\over6!}C _6
B^6-\ldots\},\label{ngw}\ee then the Schr\"odinger equation
becomes

\bq E/V&=&{1\over2}\int {d^2p\over4\pi^2}p^2C (p)\nonumber\\
0&=&-C (p)^2p^2+\hf C _4(p)p^2+1\nonumber\\
\vdots\label{ng}\eq In (\ref{ng}) $C _4(p)$ is the Fourier
transform of $C _4(x)=\int d^2yC _4(0,y,y,x)$, and we have assumed
translational invariance. The expression for the vacuum energy
density is unchanged, while the order $B^2$ part of the
Schr\"odinger equation acquires a contribution from the four-point
function. The vertical dots indicate the existence of further
equations at order $B^4$ and so on that will not concern us much
for the moment.

In addition to the equations (\ref{ng}), the vacuum wavefunctional
must satisfy normalisability conditions $C \ge0$, $C _4\ge0$, etc.
If we minimise the vacuum energy density subject to these
normalisability conditions we get $C (p)=1/p$ as before, while $C
_4$ and all other n-point functions vanish. Note that the energy
minimisation determines the contribution of the 4-point function
to the 2-point function, allowing us to solve (\ref{ng}) order by
order in $B^2$. This is a general feature of our methods.

\section{Compact QED}

The compact theory differs from the non-compact one by the
presence of Dirac monopoles, which play the role of instantons in
three space-time dimensions. As usual, we can represent the vacuum
as a sum over instanton sectors

\be\Psi[B]=\sum_n\Psi_n[B], \ee where $\Psi_n$ represents the
vacuum in the n-monopole sector. Equivalently, we can require the
vacuum to be invariant under the action of the vortex creation
operator

\be V(x)=\exp\left\{{i\over g}\int
d^2y{\epsilon_{ij}(x-y)_j\over(x-y)^2}E_i(y)\right\}, \ee so that
the vacuum wave-functional is a superposition of wave-functionals
of the non-compact theory with an arbitrary number of vortices and
anti-vortices at any spatial point \cite{11}:

\be \tilde\Psi[B]={1\over
n_+!n_-!}\sum_{n_+,n_-=0}^\infty\prod_{\alpha=1}^{n_+}\int
d^2x_\alpha \Lambda^2 V(x_\alpha)\prod_{\beta=1}^{n_-}\int
d^2x_\beta \Lambda^2V^*(x_\beta)\Psi[B]. \label{dg}\ee We have
introduced an ultraviolet cutoff $\Lambda$. If the theory is
regarded as the unbroken sector of the Georgi-Glashow model, then
$\Lambda$ is related to the charged vector boson mass $M_W$; the
details depend on the UV dynamics \cite{17}. The reason for the
aforementioned equivalence is as follows: inner-products of the
form (\ref{ex}) are regularised by point-splitting, and monopoles
that live on the quantisation surface have Dirac lines that can
intersect just one of the copies of the wave-functional in the
inner-product, inducing a vortex, or neither. Thus the vacuum,
summed over monopole sectors, is invariant under multiplication by
the vortex operator.

The vortex/monopole is represented on the wave-functional by a
shift of ${2\pi\over g}\delta^2(x)$ in the magnetic field $B(x)$.
Introducing the monopole density $\rho(x)={2\pi\over
g}\sum_{\alpha,\beta}(\delta^2(x-x_\alpha)-\delta^2(x-x_\beta))$,
the sum over monopoles in the non-Gaussian wave-functional
(\ref{ngw}) gives

\be\tilde\Psi[B]=Z^{-1}\sum_\rho\exp\{-\hf (B+\rho)C
(B+\rho)-{1\over4!}C _4 (B+\rho)^4-{1\over6!}C _6
(B+\rho)^6-\ldots\}. \label{wf}\ee We have normalised the
wave-functional to $\tilde\Psi[0]=1$ with the factor
$Z=\sum_\rho\Psi[\rho]$. The vacuum energy density is then

\be E/V={1\over2}Z^{-1}\sum_\rho\int{d^2p\over4\pi^2}\left[p^2C
(p)+[\hf p^2\rho^2C _4(p)-p^2(\rho C
(p))^2\}]+O(\rho^4)\right]\Psi[\rho], \label{ed}\ee where $\rho C
(x)=\int d^2y \rho(y)C (x-y)$, etc.

Expectation values of the monopole density are often calculated in
the effective low energy theory, which takes the form of a
sine-Gordon theory \cite{11,17}. They can also be calculated
directly by expanding the wave-functional in monopole sectors;
thus

\be Z=\sum_\rho\Psi[\rho]=1+2Vz+\ldots\ee where
$z=\Lambda^2\exp\{-{2\pi^2\over g^2}
C(0)-{2\pi^4\over3g^4}C_4(0,0,0,0)-\ldots\}$ is the monopole
fugacity, and we have included only the dominant one
(anti-)monopole sector explicitly. Note that the expression for
the monopole fugacity is to be understood in a UV regularised
sense, so that $C(0)\sim C(1/\Lambda)$, etc. It can be shown that
all volume factors cancel in the expression (\ref{wf})
(logarithmic divergences appear in the wave-functional at higher
orders, and can be removed by normal-ordering). This kind of
structure is familiar in mass perturbation theory \cite{18,19}.

To first order in $z$ we have

\be
\langle\rho(p_1)\ldots\rho(p_n)\rangle=Z^{-1}\sum_\rho\rho(p_1)\ldots\rho(p_n)\Psi[\rho]={8\pi^2\over
g^2}z\delta^2(p_1+\ldots+p_n). \ee Now to this order the
expression (\ref{ed}) for the energy density can be simplified by
using the equations (\ref{ng}). We do not need to consider $O(z)$
corrections to these equations, as they will lead to $O(z^2)$
corrections to the vacuum energy. The order $\rho^2$ terms in
(\ref{ed}) give an order $z$ term, and the order $\rho^4$ and
higher terms cancel, giving

\be E/V={1\over2}Z^{-1}\sum_\rho\int{d^2p\over4\pi^2}\left[p^2C
(p)-{8\pi^2\over g^2}z\right]\Psi[\rho].\label{ed2}\ee This is the
same result we would have obtained if we had assumed that the
wave-functional tends smoothly to the wave-functional of the
non-compact theory as $z\rightarrow0$, so that at order 1,
$C(p)=1/p$ and $C_4(p)=0$, etc.

To proceed further it is useful to expand the logarithm of the
wave-functional in powers of the magnetic field. So we write

\bq \ln\Psi&=&\ln\left(e^{-\hf
BCB-{1\over4!}C_4B^4-\ldots}\right.\nonumber\\&&\times\left.\left(1+z\int
d^2x_0\sum_\pm e^{\pm{2\pi\over g}\int
d^2xC(x_0-x)B(x)+{2\pi^2\over g^2}\int d^2xd^2y
C_4(x_0,x_0,x,y)B(x)B(y)+\ldots}\right)\right)\nonumber\\
&=&-\hf BCB +z{4\pi^2\over g^2}\left[C^2B^2-\hf
C_4B^2\right]+\ldots\nonumber\\&=&-\hf BC^{eff}B +{1\over4!}C_4^{
eff}B^4+O(B^6)+O(z^2).\label{nc}\eq Here the sum over $\pm$
corresponds to the one-monopole/one anti-monopole sectors and we
have exactly the same simplification that we observed in the
vacuum energy, giving a shift $C(p)\rightarrow
C^{eff}(p)=C(p)-{8\pi^2z\over g^2p^2}$. For higher orders in $B$
we find

\be
C_{2n}^{eff}(p_1,\ldots,p_{2n})=C_{2n}-{\delta(p_1+\ldots+p_{2n})\over
p_1\ldots p_{2n}}{2(2\pi)^{2n}z\over (2n-2)!g^{2n}}.
\label{shift}\ee Writing the equations (\ref{ng}) in terms of
$C^{eff}$, $C_4^{eff}$, etc. we have (to first order in the
monopole fugacity) the same equations and normalisability
conditions as before, and hence the same solution
$C^{eff}(p)=1/p$, $C_{2n}^{eff}=0$ for $n>1$. Thus $C(p)={1\over
p}+{8\pi^2z\over g^2p^2}$, etc.

Our failure to see mass generation at this order does not come as
so much of a surprise if we re-examine the variational Gaussian
calculation of \cite{11}. There the order $z\sim m^2$
contributions from the electric and magnetic energy densities
cancel against one another. To see the mass generation in our
formalism, we will have to extend our analysis to order $z^2$.

\section{The second-order calculation}

If we include two (anti-)monopole configurations in the
calculations of the last section, we find the two monopole
contribution

\be \langle\rho(x)\rho(y)\rangle_2=z^2\left(\pm e^{\pm
C(x-y)}+\delta^2(x-y)\int_{w>L} d^2w\left(e^{\pm
C(w)}-1\right)\right). \ee Here the sign is $\pm$ as the monopoles
have equal or opposite charge, and in the second term we have a
partial cancellation of the infrared divergence due to the
normalisation factor $Z$, though a logarithmic divergence remains.
The monopole density in the vacuum is low for $z\ll1$ so that
(\ref{dg}) is the partition function for a dilute gas, in which
monopoles are assumed to be widely separated, hence the cutoff in
the integral.

Putting this result into (\ref{nc}) we find the order $z^2$
contribution to $C^{eff}(x-y)$:

\bq &&8\pi^2{z^2\over g^2}\left\{\left[\int_{|x_1-x_2|>L}
d^2x_1d^2x_2{1\over |x-x_1|}{1\over |x_2-y|}\sinh\left({1\over
g^2|x_1-x_2|}\right)\right]\right.\nonumber\\&&\left.\qquad+\delta^2(x-y)\int_{w>L}
d^2w\left(\cosh\left({1\over
g^2w}\right)-1\right)\right\}.\label{oz2}\eq In the dilute gas
approximation we can take $\sinh\left({1\over
g^2|x_1-x_2|}\right)\approx {1\over g^2|x_1-x_2|}$ and evaluate
the first integral in (\ref{oz2}) as ${|x-y|\over g^2}$. This
corresponds to a mass term in the two-point function; the puzzle
is that the mass $m^2\sim {z^2\over g^4}$ appears to be of order
$z^2$, whereas we know from \cite{17} that the effective photon
mass is of order $z$.

Again, the problem is that the topological expansion makes it
difficult to read the physical mass off from the wave-functional
directly. What we need to determine the physical mass is a vacuum
expectation value like $\langle B(x)B(y)\rangle$, which can be
read off from the functional Fourier transform of the non-Gaussian
wave-functional that we have constructed.

To obtain the functional Fourier transform we treat the four-point
and higher functions as perturbations of the quadratic
wave-functional. The Schr\"odinger equation again has the solution
$C^{eff}(p)=1/p$, $C_{2n}^{eff}=0$ for $n>1$, and we can write the
wave-functional in terms of the field $A_i$ (with vector indices
absorbed into the matrix notation) as

\be \Psi[A_i]=\exp\{-\hf A\tilde C  A-{1\over4!}\tilde C_4
A^4-{1\over6!}\tilde C _6 A^6-\ldots\},\ee where for example
$\tilde C(p)={\cal P}p^2C(p)$ with ${\cal P}$ a transverse
projector. For our purposes we can now forget about the
topological expansion, since VEV's of the magnetic field operator
are unaffected by the presence of monopoles \cite{11}. We are
interested in the quadratic part of the Fourier transformed
wave-functional, from which we can read off the physical mass.

The latter receives contributions from tadpole diagrams like $\int
d^2x_1\ldots d^2x_{2n}\tilde C_{2n}(x_1,\ldots,x_{2n})$ $\tilde
C^{-1}(x-x_1)\tilde C^{-1}(x_2-x_3)\ldots\tilde C^{-1}(x_{2n}-y)$,
etc. which, when summed, give the final result for the photon mass

\be m^2\sim{z^2\over g^4}\exp\left(-{\pi\Lambda\over
2g^2}\right)={\Lambda^2\over g^4}z. \ee This result agrees with
\cite{11,17}, and is of first order in the monopole fugacity, as
expected. Interestingly, the dynamical mass generation is given
here by the condensation in the vacuum of a monopole/anti-monopole
pair.

\section{Conclusions}

For the case of compact QED in (2+1) dimensions we have shown how
the mass generation first demonstrated by Polyakov \cite{17} and
reproduced in the Hamiltonian formalism in \cite{11} by means of a
gauge-projected Gaussian ansatz, can be seen in the full
non-Gaussian wave-functional. There is a variational principle
implicit in any solution of the Schr\"odinger equation, and our
method provides a link between variational methods based on a
Gaussian ansatz, and attempts to solve the functional
Schr\"odinger equation exactly in some non-perturbative expansion.

In our analysis the variational principle and normalisability
conditions on the wave-functional played a role, but the mass
generation was seen to be a feature of the effective monopole
dynamics. In general, explicit expansion around topological
solutions may be the best way to tackle a systematic Hamiltonian
analysis of non-abelian gauge theories.

The extension of these functional Schr\"odinger methods to
fermions and superfields is straightforward. The study of
supersymmetric theories along these lines is strongly suggested by
the non-renormalisation theorems afforded by such theories; these
simplify the quantisation considerably, as well as facilitating
the interpolation between different perturbative regimes.

\end{document}